\documentclass[12pt]{article}
\usepackage[a4paper, total={17cm, 25cm}]{geometry}
\usepackage[utf8]{inputenc}
\usepackage{authblk}
\usepackage{graphicx}
\usepackage[dvipsnames]{xcolor}
\usepackage{amsmath}
\usepackage{amssymb}
\usepackage{amsfonts}
\usepackage{dsfont}
\usepackage{diagbox}
\usepackage{upgreek}
\usepackage{hyperref}
\usepackage{cite}
\usepackage{siunitx}
\usepackage[normalem]{ulem}

\author[1]{Erik Olsén}
\author[2]{Berenice García Rodríguez}
\author[2]{Fredrik Skärberg}
\author[1]{Petteri Parkkila}
\author[2]{Giovanni Volpe}
\author[1]{Fredrik H{\"o}{\"o}k}
\author[2]{Daniel Midtvedt}

\affil[1]{{\small Department of Physics, Chalmers University of Technology, Gothenburg, Sweden}}
\affil[2]{{\small Department of Physics, University of Gothenburg, Gothenburg, Sweden}}

\title{Dual-angle interferometric scattering microscopy for optical multiparametric particle characterization}

\begin{document}

\maketitle

\begin{abstract}
\noindent Traditional single-nanoparticle sizing using optical microscopy techniques assesses size \textit{via} the diffusion constant, which requires suspended particles in a medium of known viscosity.
However, these assumptions are typically not fulfilled in complex natural sample environments.
Here, we introduce dual-angle interferometric scattering microscopy (DAISY), enabling optical quantification of both size and polarizability of individual nanoparticles without requiring \textit{a priori} information regarding the surrounding media or super-resolution imaging.
DAISY achieves this by combining the information contained in concurrently measured forward and backward scattering images through twilight off-axis holography and interferometric scattering (iSCAT).
Going beyond particle size and polarizability, single-particle morphology can be deduced from the fact that hydrodynamic radius relates to the outer particle radius while the scattering-based size estimate depends on the internal mass distribution of the particles.
We demonstrate this by optically differentiating biomolecular fractal aggregates from spherical particles in fetal bovine serum at the single particle level.
\end{abstract}


\vspace{.2in}

\noindent Single-nanoparticle characterization in terms of size, shape, and composition in complex biological environments is a critical challenge within several research areas, including drug delivery\cite{jindal2017effect}, diagnostics\cite{chen2016nanochemistry} and nanosafety\cite{zielinska2020nanotoxicology}.
Optical microscopy is in many cases the tool of choice for studying individual biological nanoparticles due to its high throughput and biological compatibility\cite{priest2021scattering}. 
However, although nanoparticles as small as individual proteins can be detected using label-free optical scattering microscopy\cite{kashkanova2021precision,vspavckova2022label,young2018quantitative,dahmardeh2023self}, multiparametric characterization of individual nanoparticles in terms of properties such as size, refractive index and morphology remains a challenge.

Since nanoparticles are smaller than the spatial resolution of optical scattering microscopy, it is difficult to estimate their size from direct observation in a microscopy image.
Instead, size is typically estimated indirectly by tracking their position over time, estimating their diffusivity from their trajectories, and finally using the Stokes-Einstein relation to relate single-particle diffusivity to the particle size\cite{bian2016111}.
However, this requires that the nanoparticles are freely diffusing in a medium with known viscosity.
This imposes critical limitations when the analysis is carried out in the natural environment of the particles (\textit{in situ}).
For example, the viscoelastic properties of biological environments are typically complex and may exhibit spatial variations\cite{efremov2020measuring}, which severely limits the applicability of particle sizing using diffusivity.

Quantitative \textit{in situ} particle characterization using optical microscopy must instead directly relate the optical scattering of individual particles to their physical properties.
The scattering amplitude depends on the particle polarizability, defined as
\begin{equation}
    \alpha\equiv 3V\frac{n_{\text{p}}^2-n_{\text{m}}^2}{n_{\text{p}}^2+2n_{\text{m}}^2},
    \label{eq:pol1}
\end{equation}
where $n_{\text{m}}$ and $n_{\text{p}}$ are the media and particle refractive indices, respectively, and $V$ is the particle volume. 
In the limit of small refractive index differences between the particle and the media, the polarizability relates to particle properties as $\alpha\approx \left(2/n_{\rm m}\right) V\Delta n$, where $\Delta n=\left(n_{\rm p}-n_{\rm m}\right)$.
Hence, the polarizability is to a first approximation proportional to the difference in refractive index between particle and medium. 
Combined with that $\Delta n$ scales essentially linearly with molecular concentration\cite{zangle2014live}, polarizability information can be used to estimate particle mass, as demonstrated for both single biomolecules and living cells\cite{young2018quantitative,zangle2014live}.
However, since Eq.~\eqref{eq:pol1} depends on both particle volume and $\Delta n$, the scattering amplitude alone is insufficient to characterize both quantities at once.

In addition to the scattering amplitude, the angular distribution of light scattering also contains information about particle size and morphology\cite{bohren2008absorption}. 
This forms the basis of particle characterization using multi-angle light scattering (MALS)\cite{wyatt1998submicrometer} and scattering-based flow cytometry\cite{van2018absolute}. 
In the context of microscopy, images of scattering patterns have been employed for simultaneous estimation of size and refractive index of particles with diameters down to about half the wavelength of light\cite{midtvedt2021fast,altman2023machine}.
This lower size limit originates from the difficulty of accurately relating a measured scattering image to particle size for particles near the diffraction limit.

\begin{figure}[!ht]
\centering
\includegraphics[width=1\textwidth]{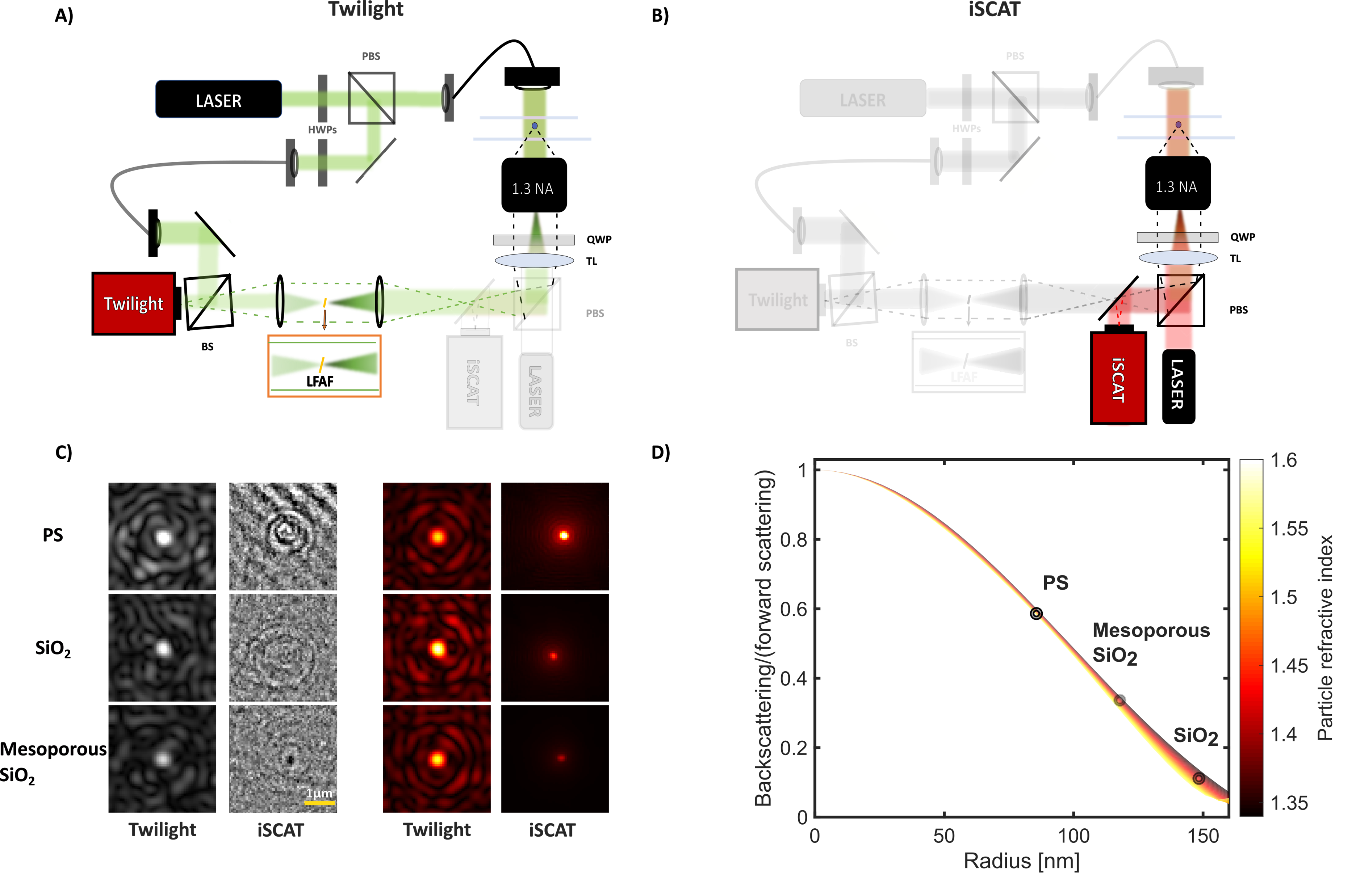}
\caption{\textbf{DAISY working principle}. {\bf A, B} Optical setup to enable simultaneous twilight off-axis holography (highlighted in {\bf A}) and interferometric scattering (iSCAT) (highlighted in {\bf B}) measurements. Using two different wavelengths for holography and iSCAT, the two signals are separated by a dichromatic mirror and directed to two cameras. The low frequency attenuation filter (LFAF) reduces the amplitude of the unscattered light of the sample beam while having a negligible effect on the particle signal (Supporting information, Section 1.7). The reduction of the unscattered light is highlighted in the zoomed-in inset. The LFAF is slightly tilted to direct the reflected light away from the optical axis. BS: beam splitter, OBJ: objective, TL: tube lens, QWPs: Quarter-wave plates, and HWPs: Half-wave plates. {\bf C} After detection, the images of particle scattering patterns are postprocessed before the signals are quantified. The two left columns are the scattering patterns after background subtraction, and the two right columns are the same particles but further postprocessed, where the twilight images are the average particle signal along a trace and the iSCAT image has been processed using a U-Net. See Supporting Information Section 1.5 for more information about the postprocessing. PS: polystyrene and SiO$_2$: silica. {\bf D} The ratio between the amplitudes of backscattered and forward scattered optical fields as a function of particle radius for spherical particles in water. The gradient color scale encodes the dependence of the scattering ratio on the particle refractive index from 1.35 to 1.60. The three circles indicate the size and refractive index of the particles in {\bf C}.}
\label{fig:iSCAT}
\end{figure}

In this work, we introduce dual-angle interferometric scattering microscopy (DAISY), which offers simultaneous quantification of both size and polarizability (and hence also $\Delta n$) of individual particles beyond the limits set by diffraction (around half the wavelength of light) without requiring precise information about the surrounding medium. 
DAISY exploits the fact that the forward and backward scattered optical signals of individual particles scale differently with size and refractive index, which enables characterization of particle size and polarizability, given that the optical signals are measured simultaneously (Figure \ref{fig:iSCAT}).
The forward scattering image is measured using twilight off-axis holography (Figure \ref{fig:iSCAT}A), which quantifies the complex-valued optical field\cite{olsen2023virus}, while the backscattering image is measured using interferometric scattering (iSCAT) microscopy (Figure \ref{fig:iSCAT}B), probing the interference between the backscattered particle signal and a coherent background signal\cite{priest2021scattering} (Supporting Information, Section 1.3).
The twilight holography and iSCAT images (Figure \ref{fig:iSCAT}C) are processed using standard algorithms for off-axis holography\cite{kim2010principles} and a U-Net trained to generate focused particle images\cite{midtvedt2021quantitative} where the signal is proportional to the scattering amplitude in the backward direction (Supporting Information, Section 1.5), respectively.
After that, the particle signal in each image is estimated by taking the spatial integral of the post-processed images using a 2D Gaussian fit.

On the one hand, the optical field in the forward direction is proportional to the polarizability as (Supporting information, Section 2.3)\cite{khadir2020full}
\begin{equation}
   \alpha=\frac{\lambda_{\text{0}}}{n_{\text{m}}\pi}\iint \Im(E_{\text{p}})dA
    \label{eq:alpha},
\end{equation}
where $\lambda_{\text{0}}$ is the illumination wavelength in vacuum and $\Im(E_{\text{p}})$ is the imaginary part of the scattered field in the forward direction. 
On the other hand, the optical signal in the backward direction is proportional to the product between polarizability and self-interference effects from different scattering elements within the particle, where the interference effects can be described by the optical form factor $f$ given by Rayleigh-Debye-Gans (RDG) theory\cite{bohren2008absorption}.
Therefore, the ratio between the scattering signal in the backward and forward directions is proportional to the optical form factor in the limits within which RDG theory accurately describes the particle signal (i.e., $\vert n_p/n_m -1\vert \ll 1$ and $\vert n_p/n_m -1\vert kR \ll 1$, where $k=2\pi n_m/\lambda_0$ and $R$ is the particle radius).

Using RDG theory, the optical form factor of an isotropic particle is\cite{bohren2008absorption}
\begin{equation}
    f(q;\rho) = \frac{1}{V}\int dr r^2 \rho(r) \frac{\sin(qr)}{qr},
    \label{eq:form}
\end{equation}
where $q=(4\pi/\lambda_0) n_{\rm m}\sin(\theta/2)$, $\theta$ is the angular difference between the incoming and scattered light, and $\rho(r)$ is the spatial distribution of scattering elements within the particle. 
Noticeably, the optical form factor is independent of the particle refractive index, which in MALS is used to relate the optical scattering to particle size\cite{wyatt1998submicrometer}.
Moreover, the optical form factor only depends on the medium refractive index through the dependence of the $q$-number on $n_{\rm m}$, which enters as a product with particle radius.
Since the refractive index of biological environments is close to that of water (for example, the refractive index of the cytoplasm of a cell is only a few percent higher than that of water\cite{gul2021cell}), this suggests that particle size can be inferred from the optical form factor with a precision within a few percent without detailed information about the surrounding media (Supporting information, Section 2.4).

The size range for which RDG theory accurately describes the optical scattering depends on the refractive index difference between the particle and the surrounding media\cite{bohren2008absorption}.
In the case of spherical particles, Mie theory can be employed beyond the limitations of RDG.
This is illustrated in Figure \ref{fig:iSCAT}D by plotting the ratio between the amplitudes of backscattered and forward scattered optical fields as a function of particle radius for spherical particles in water.
In the case of homogeneous spheres with a known refractive index and radius less than 170~nm, the scattering ratio is uniquely related to particle size. 
Noticeably, this upper size limit coincides with the lower size limit for which the radius and polarizability can be determined directly from an off-axis holography image\cite{midtvedt2021fast}.
When the particle refractive index is unknown, the estimated spread in particle radius based on the scattering ratio is approximately 3~nm for particles with a radius smaller than 100~nm, even if the particle refractive index varies between 1.35 and 1.60 (Figure \ref{fig:iSCAT}D); and even for particles with a radius of 150 nm, the corresponding uncertainty does not exceed 10~nm.
Thus, DAISY has the potential of extending the lower size limit image-based particle sizing, with the precision of the size estimate being dependent on the available refractive index information about the particle.

We define the DAISY radius (denoted by $r_{\rm DAISY}$) as the smallest radius of a homogeneous sphere suspended in water displaying the same backward--forward scattering ratio and polarizability, where the inclusion of polarizability improves the accuracy when relating the scattering ratio to size.
To extend the particle sizing beyond the limitations of RDG theory, we here introduce the generalized form factor $\Tilde{f}$ as the theoretical scattering ratio obtained using Mie calculations, normalized such that $\Tilde{f}(R=0)=1$ to make it similar to the optical form factor.
The scattering ratio measured in DAISY can be related to the generalized form factor as
\begin{equation}
   \frac{\text{iSCAT}}{|\text{twilight}|} = C \frac{|E_p(\text{backward})|}{|E_p(\text{forward})|} \equiv C \Tilde{f}(q_b;\rho,\alpha), 
    \label{eq:ratio}
\end{equation}
where $C$ is a calibration constant obtained by comparing reference measurements of known particles and $q_{\rm b}$ is the effective wave number of the iSCAT measurement (Supporting information, Section 1.8). 
The generalized form factor is in turn, related to particle size.

\begin{figure}[!ht]
\centering
\includegraphics[width=0.7\textwidth,trim= 8 2 2 12,clip]{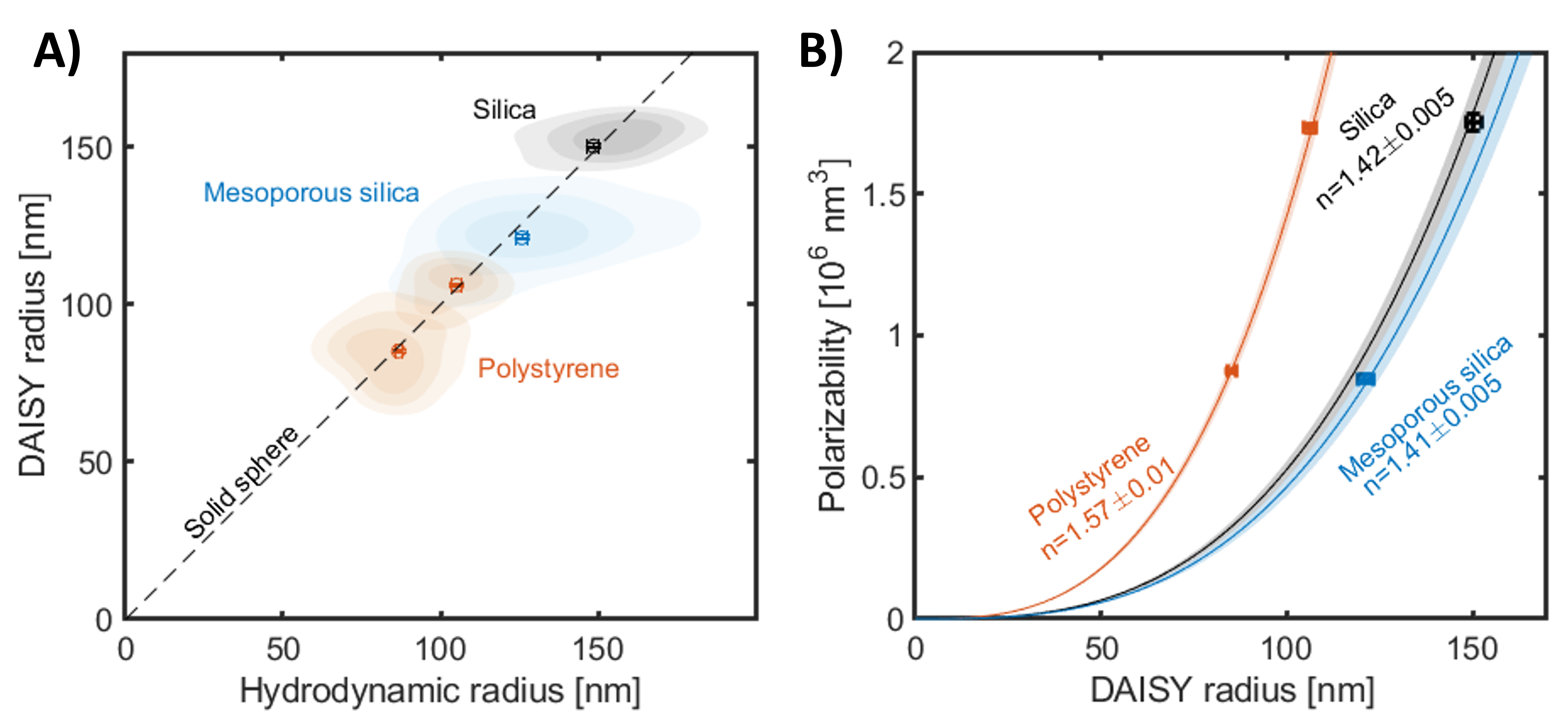}
\caption{\textbf{Evaluation of the DAISY radius using the simultaneously obtained hydrodynamic radius as reference}. {\bf A} The DAISY radius with the simultaneously obtained hydrodynamic radius for two polystyrene samples, one silica sample, and one mesoporous silica sample, suspended in water. All measured particles follow a one-to-one relationship with the hydrodynamic radius, where the median difference between the two size estimates is less than 5\% for all particle populations. The shaded areas correspond to a contour plot of the DAISY radius and the hydrodynamic radius, where the median value is the point in the plots. {\bf B} Refractive index estimates using the polarizability information from twilight holography and the DAISY radius in {\bf A}. The estimated refractive indices are  $1.57\pm0.01$ for polystyrene, $1.42\pm0.005$ for silica, and $1.41\pm0.005$ for mesoporous silica, where the solid line is the estimated refractive index value and shaded region corresponds to the uncertainty in the refractive index estimate.}
\label{fig:size}
\end{figure}

To validate this hypothesis, spherical particles of different sizes and refractive indices were measured under flow in a microfluidic channel when suspended in water (Supporting information, Section 1.4). 
Specifically, two polystyrene samples ($R=85\pm13$~nm and $R=105\pm23$~nm), one silica sample ($R=150\pm28$~nm), and one mesoporous silica sample ($R=120\pm29$~nm) were measured, where the $\pm$ is the standard deviation of the distribution measured using darkfield nanoparticle tracking analysis (NTA) (Supporting information, Figure S6).
Since these particles were measured freely diffusing in water, the assumptions underlying diffusivity-based particle sizing are fulfilled, enabling simultaneous determination of both the hydrodynamic radius and the DAISY radius. 

The estimated DAISY radius and hydrodynamic radius exhibit a one-to-one correspondence, with a deviation in median size of less than 5\% for all particles, as shown in Figure \ref{fig:size}A. 
Moreover, the distribution widths of the DAISY radius are consistently similar to or smaller than those of the hydrodynamic radius. 
This suggests that the DAISY radius estimation is more precise than the hydrodynamic radius when the same track length is used (Supporting information, Section 1.8). Consequently, DAISY effectively estimates particle sizes below the diffraction limit in microscopy images without relying on super-resolution imaging or detailed information about the experimental point spread function.

By using the simultaneously quantified DAISY radius and polarizability, refractive indices for the measured polystyrene, silica, and mesoporous particles were determined to be $1.57\pm0.01$, $1.42\pm0.005$, and $1.41\pm0.005$, respectively, as depicted in Figure \ref{fig:size}B. 
Notably, all of these values are within a 0.02 refractive index difference from prior estimates \cite{nikolov2000optical,ma2003determination,van2014refractive,kashkanova2021precision,odete2020role}. 
This confirms that DAISY enables accurate image-based nanoparticle characterization in terms of size and polarizability without reliance on the Stokes-Einstein relation.

\begin{figure}
\centering
\includegraphics[width=8cm]{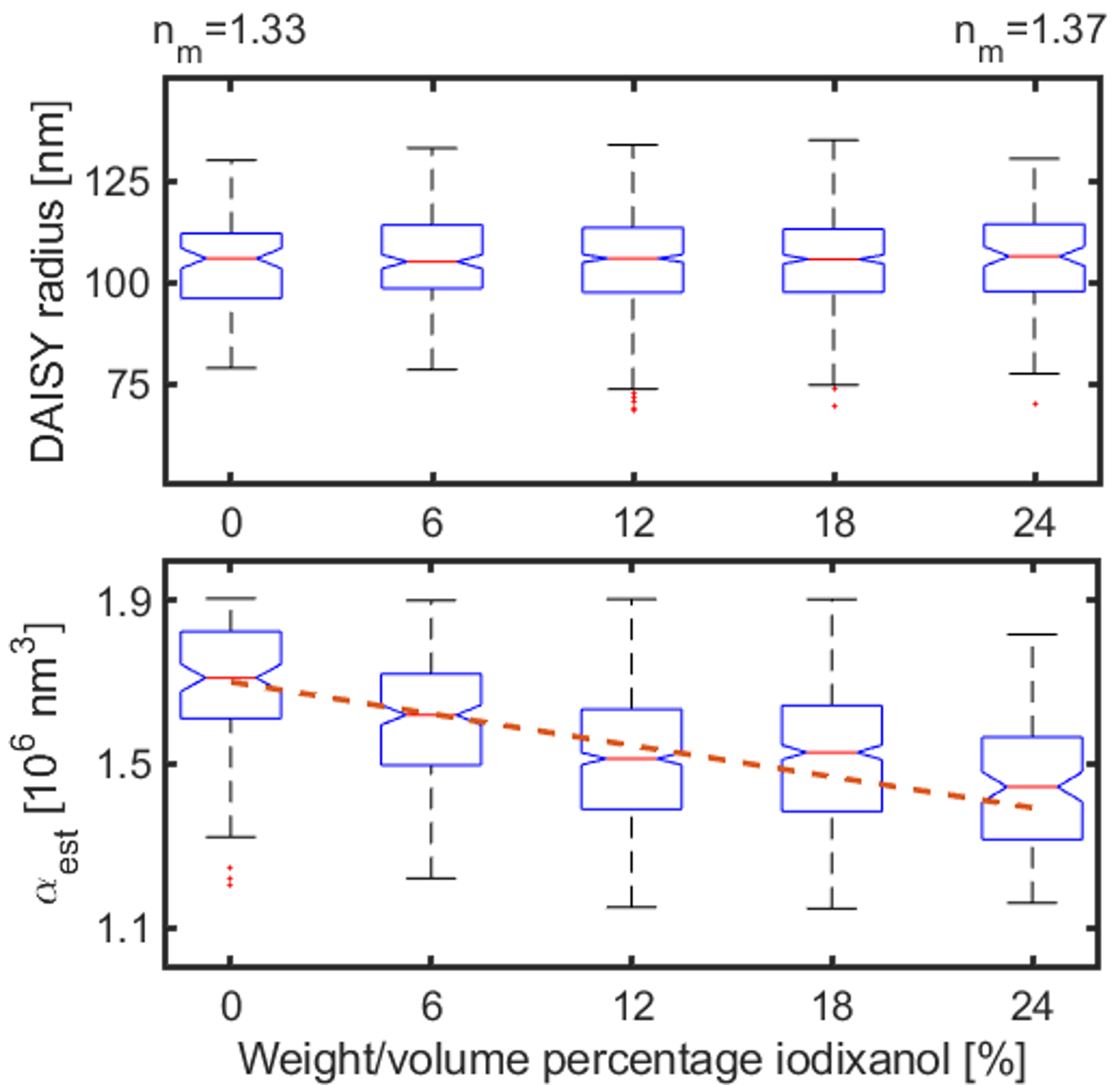}
\caption{\textbf{Evaluation of DAISY radius and polarizability determination in different media}. Box plots of the DAISY radius and polarizability as a function of the water-iodixanol concentration for 105~nm radius polystyrene spheres. The DAISY radius remains the same for all different media, whereas the polarizability decreases as the refractive index difference to the surrounding media decreases. The DAISY radius and the effective polarizability are here estimated assuming that the surrounding refractive index has a refractive index as water. The dashed line is the theoretical polarizability for a 105~nm radius polystyrene sphere as a function of the surrounding refractive index.}
\label{fig:pol}
\end{figure}

To verify that the DAISY radius is indeed insensitive to the precise information regarding the surrounding media, we measured one particle sample (polystyrene spheres, modal radius $105$~nm) in aqueous environments with varying amounts of water and iodixanol, thereby varying the refractive index of the environment (Figure \ref{fig:pol}).
Even though DAISY radius and polarizability are estimated as if the particles are in water, the median DAISY radius remains close to the nominal value of 105~nm and varies by less than 2~nm when the surrounding refractive index is changed from 1.335 to 1.37 (corresponding to a variation in iodixanol concentration between 0\% and 24\%\cite{boothe2017tunable}).
This low spread in size estimation is consistent with the observation made in connection with Eq. \eqref{eq:form} regarding that the error in the estimated DAISY-radius is bounded by the error in the $q$-number used when relating the estimated generalized form factor to the DAISY radius (Supporting Information, Section 2.4).
Given the width of the DAISY radius distribution, the spread in median DAISY radius estimates most likely originates from the statistical uncertainty in estimating the median radius rather than any systematic media refractive index dependence.
Moreover, the particle polarizability estimation decreases with increasing media refractive index (Figure \ref{fig:pol}B), as expected from Eq. \eqref{eq:alpha} since the polarizability is dependent on the refractive index difference between the particle and media. 
These results indicate that the particle size estimation offered by DAISY remains accurate as long as the relative error in the $q$-number is small ($\ll 1$), which, if combined with the polarizability information, enables estimation of the refractive index difference between the particle and the surrounding media.

Note that the DAISY radius is complementary to the hydrodynamic radius, and these two size estimates coincide in the case of homogeneous spheres.
To a first approximation, the hydrodynamic radius reflects the physical boundary of the particle, whereas from Eq. \eqref{eq:form} the DAISY radius also reflects the interior mass distribution of the particle. 
Thus, the relation between the DAISY and hydrodynamic radius can provide information about the spatial distribution of mass within the particle and its morphology.

\begin{figure}[!ht]
\centering
\includegraphics[width=17cm,trim= 2 2 2 15,clip]{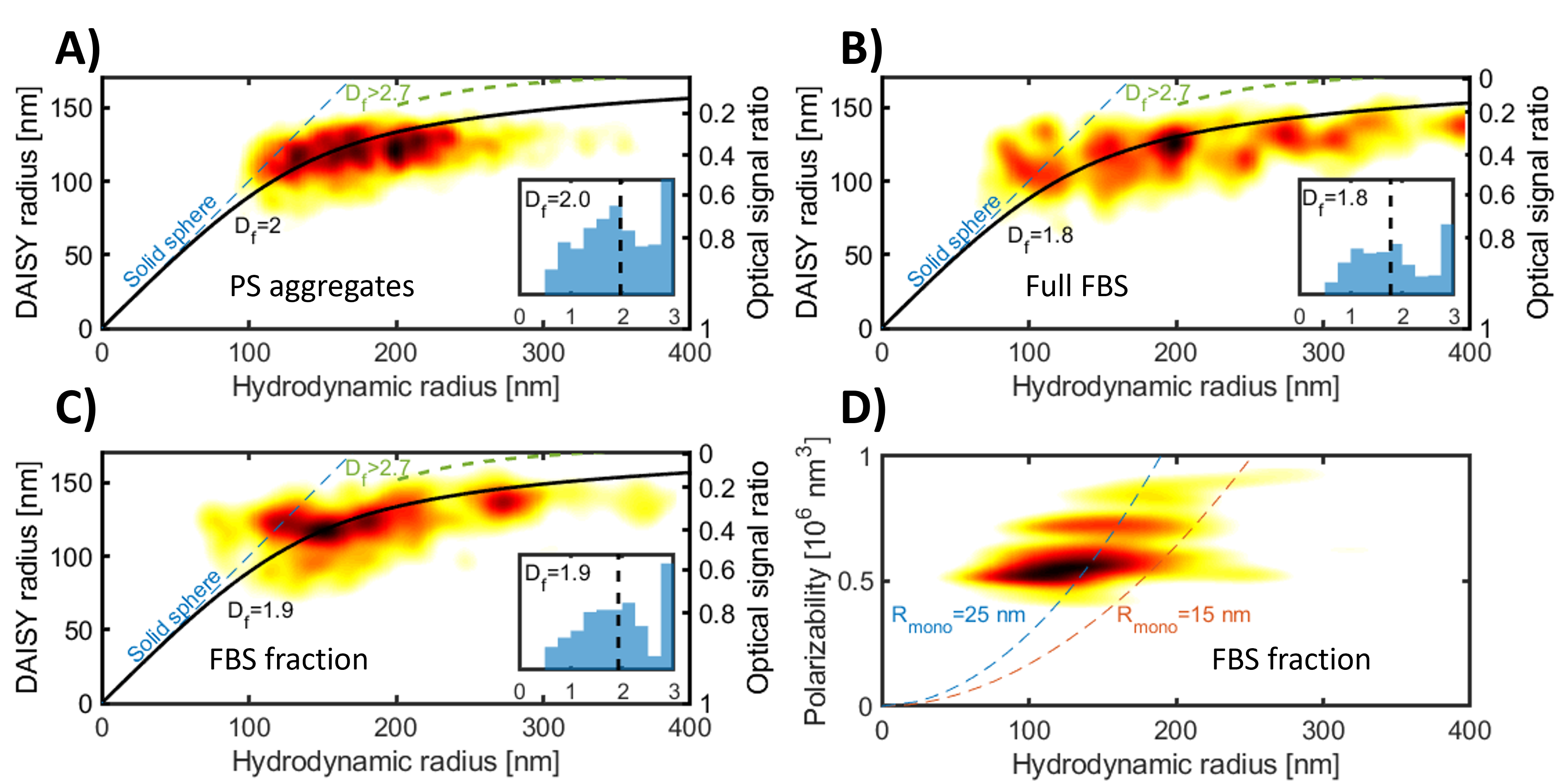}
\caption{\textbf{DAISY radius and hydrodynamic radius to distinguish particle aggregates from solid spheres}. {\bf A} DAISY radius as a function of hydrodynamic radius for salt-induced aggregation of 35~nm radius polystyrene particles. From the comparison with the theoretical lines, the DAISY-hydrodynamic radius relation agrees with that of fractal aggregates with a fractal dimension of around 2.0. {\bf B-C} The DAISY radius and hydrodynamics radius for freeze-thawed induced aggregates of fetal bovine serum (FBS), both {\bf B} in full serum and {\bf C} after size-exclusion chromatography. In {\bf A}-{\bf C}, the curved lines correspond to the theoretical relation for fractal aggregates, and the straight dashed line is the expected scaling for a solid sphere. The green dashed lines correspond to a fractal dimension of 2.7, which separates true aggregate detections from that of spheres with a hydrodynamic radius larger than 200 nm. The insets are single-particle fractal dimension histograms, where the dashed line corresponds to the median fractal dimension value. The number of particle detections in FBS corresponds to a concentration of around $10^8$/ml. {\bf D} Polarizability as a function of hydrodynamic radius for the data in  {\bf C}, where the dashed lines are the expected scaling for aggregates with a fractal dimension of 1.9 and monomer refractive index of 1.53.}
\label{fig:ratio} 
\end{figure}

To evaluate the potential of DAISY for estimating particle morphology, we formed aggregates containing 35~nm radius polystyrene spheres via salt-induced aggregation.
The detected aggregates deviate from the one-to-one relation between the DAISY radius and the hydrodynamic radius previously observed for homogeneous spheres (Figure \ref{fig:ratio}A), indicating that the morphology of the aggregates is different from that of homogeneous spheres.
To develop this observation into a quantitative analysis, we first note that particle aggregates are generally well described as fractal aggregates\cite{lattuada2003hydrodynamic,sorensen2001light}.
Treating the aggregates as spherical units, their mass density is a decaying function of the radial distance from the aggregate center, $n(r)\sim r^{D_{\rm f}/3-1}$, where $D_{\rm f}$ is the fractal dimension and is expected to fall within the range 1.5-2.3 for particle aggregates\cite{schaefer1984fractal,carpineti1990salt,midtvedt2021fast}.
An explicit relation between the DAISY radius and overall aggregate radius, here approximated by the hydrodynamic radius, can be derived from theoretical models of fractal aggregates for which the fractal dimension is the only free parameter (Supporting Information, Section 2.5)\cite{lattuada2003hydrodynamic,sorensen2001light}.
Since the size range of an unique relation between the optical scattering ratio and particle size depends on particle morphology (Supporting Information, Section 2.6), we found that the $(r_{\rm DAISY},r_{\rm H})$-space can be subdivided into two disjoint regions for the sizes in Figure \ref{fig:ratio}A.
One of the regions encompasses spherical, non-fractal monomeric units (including homogeneous spheres), whereas the other region encompasses fractal aggregates having fractal dimensions $D_{\rm f}<2.7$. At this threshold value of the fractal dimension, the theoretical scattering ratio curves for aggregates and homogeneous spheres tangent each other, hindering a reliable separation between fractal and non-fractral structures with fractal dimensions exceeding this value (Supporting Information, Figure S7).
Furthermore, each point within the fractal aggregate region is related to a specific value of the fractal dimension, indicating that the fractal dimension of individual aggregates can be quantified based on their position in $(r_{\rm DAISY},r_{\rm H})$-space.  
We found that the salt-induced aggregates generally fall within the region in $(r_{\rm DAISY},r_{\rm H})$-space encompassing fractal aggregates, validating the analysis approach outlined above (Figure \ref{fig:ratio}A). 
The polystyrene aggregates have a population-wide median fractal dimension of $D_{\rm f}=2.0$, consistent with expectation for diffusion-limited cluster aggregation (inset to Figure \ref{fig:ratio}A)\cite{schaefer1984fractal,carpineti1990salt}.%

To investigate whether the same analysis approach can provide information about the morphology of the constituents in more complex solutions, we performed measurements of freeze-thawed fetal bovine serum (FBS) both non-treated and depleted of proteins (Supporting Information, Section 1.2). 
In addition to individual dissolved biomolecules, FBS contains biological particles such as extracellular vesicles (EVs), lipoprotein particles and protein aggregates\cite{mannerstrom2019extracellular}.
Using DAISY, we detected particles with a hydrodynamic radius between 100 and 400 nm in both ordinary FBS and FBS separated from free proteins using size-exclusion chromatography (Figure \ref{fig:ratio}B-C), at a concentration of about $10^8$/ml. 
Notably, DAISY's detection count was $10^3$ times lower than darkfield measurements (Supporting information, Figure S1), suggesting that it primarily identifies larger particles or aggregates above its detection limit.
The particle detections with a hydrodynamic radius around 150-200 nm have a fractal dimension close to $D_{\rm f}=2.0$, whereas the larger particles have a fractal dimension in the vicinity of $D_{\rm f}=1.7$.
In addition to this, a small fraction of detections with a hydrodynamic radius around 100-150~nm deviate from the fractal aggregate scaling, in particular for the FBS after size-exclusion chromatography, and instead coincide with the expected scaling for homogeneous spheres (Figure \ref{fig:ratio}C).
These detections have a polarizability of around $0.55\times 10^6$~nm$^3$, which together with a hydrodynamic radius of 125 nm corresponds to a refractive index of about 1.38 (Figure \ref{fig:ratio}D).
This value is similar to the expected values for EVs\cite{van2014refractive,gardiner2014measurement,kashkanova2022interferometric}, which, if filled with biological material, are expected to have an optical form factor similar to homogeneous spheres.
However, identification of EV surface markers is required for conclusive identification of the presence of EVs.
Nevertheless, the rich single-particle shape information using DAISY indicates that it enables analysis of sub-populations and heterogeneity within the sample, which extends the possibilities compared to previous works where particle shape is estimated on the ensemble level from the signal-size scaling\cite{kashkanova2021precision,midtvedt2019size,midtvedt2021fast}.

To gain additional insights into the nature of the fractal aggregate population of FBS, we investigated the relation between polarizability and hydrodynamic radius for FBS after size-exclusion chromatography. 
The polarizability of a fractal aggregate is directly proportional to the number of monomers $N$ in the aggregate as $\alpha=\alpha_0N$, where $\alpha_0$ is the polarizability of the monomers.
Since the hydrodynamic radius also scales with the number of monomers and the fractal dimension is known from the relation between DAISY radius and hydrodynamic radius, we can estimate the monomer polarizability.
Assuming that the monomer has a refractive index of 1.5, which is similar to lipid bilayers, proteins, and lipoprotein particles\cite{kashkanova2022interferometric,parkkila2018biophysical,zhao2011distribution}, we find that the monomer has a hydrodynamic radius around 20-30~nm (Figure \ref{fig:ratio}D). 
This value is considerably larger than individual proteins (having a typical radius of less than 10 nm)\cite{erickson2009size}.
The estimated properties of the aggregate are thus consistent with a larger monomer, with lipoproteins being a likely candidate\cite{kashkanova2022interferometric}. 
However, viral or EV monomers cannot be excluded based on this data alone\cite{rensen2001recombinant,kashkanova2022interferometric}.
It should be noted that these results do not exclude the presence of protein aggregates with smaller monomer units in FBS; it only demonstrates that the aggregates that were detected in our setup consist of monomers of this size.

In conclusion, we have introduced a versatile method for multiparametric particle characterization, namely, dual-angle interferometric microscopy (abbreviated DAISY). 
We have demonstrated the capacity of DAISY to simultaneously quantify size and polarizability of particles directly from optical scattering patterns without being limited by the diffraction limit.
Moreover, the DAISY radius shows negligible dependence on the refractive index of the surrounding media and is complementary to the hydrodynamic radius, allowing the combination of the DAISY and hydrodynamic radius to provide particle morphology estimates. 
Thus, DAISY opens up for analysis of particle morphology for freely suspended particles as well as temporally resolved \textit{in situ} monitoring of nanoparticles in biological environments.
The ability to measure in biological environments make DAISY a promising candidate for single-particle analysis of particles inside cells in terms of size and polarizability, in particular considering that single-particle analysis for particles inside cells has been previously demonstrated using both holography and iSCAT separately\cite{midtvedt2022single,sandoghdar2022confocal}.
Moreover, by analyzing the full scattering pattern instead of only estimating the integral of the particle images, both the size range of DAISY and its ability to differentiate different particle morphologies, such as analyzing asymmetric particles\cite{abdulali2022multi}, can likely be further improved.
Given DAISY's versatility and the presented characterization opportunities, we anticipate that this type of optical microscopy-based multiparametric characterization will find widespread application in many areas where nanoparticles play an important role, ranging from industrial processes to drug discovery and medical diagnostics.

\section{Acknowledgments}

This research was funded by the Swedish research council, grant number 2019-05071, the Knut and Alice Wallenberg Foundation grant number 2019-0577, and Chalmers Area of Advance Nano. 
Myfab is acknowledged for support and access to the nanofabrication laboratories at Chalmers.
We also thank Dr. Björn Agnarsson for fabricating the LFAF.

\section{Author contributions}

EO, FH, and DM conceived the method. 
EO and BGR implemented the method. 
EO and BGR evaluated the method and collected the DAISY data using an experimental setup and software developed by EO, DM, and FH. 
EO and DM performed the DAISY data analysis.
PP prepared and analyzed serum samples.
FS and DM designed and trained the neural network.
GV, FH, and DM supervised the work. 
EO and DM drafted the paper.
EO, BGR, GV, FH, and DM drafted the illustrations. 
All authors revised the paper. 

\section{Competing interests} 

The authors declare the following competing financial interest(s): E.O. and D.M. own shares in a private company (Holtra) that holds IP related to twilight microscopy method.

\section{Supplementary information}

Materials and methods, data analysis, and theoretical descriptions of the DAISY radius and the optical scattering ratio for fractal aggregates.

\bibliographystyle{ieeetr}
\bibliography{references}

\end{document}